\documentstyle[12pt]{article}
\def\gev{{\rm \, Ge\kern-0.125em V}}
\begin{document}
\begin{titlepage}
\pagestyle{empty}
\baselineskip=21pt
\rightline{hep-ph/9402233}
\rightline{UMN--TH--1233/94}
\rightline{UCSBTH--94--02}
\rightline{February 1994}
\vskip1.25in
\begin{center}
{\large{\bf
Corrections to Bino Annihilation II: One-Loop Contribution to
$\widetilde B\widetilde B \to Z^*$ }}
\end{center}
\begin{center}
\vskip 0.5in
{Toby Falk,$^1$ Richard Madden,$^2$ \\
Keith A.~Olive,$^2$  and Mark Srednicki$^1$
}\\
\vskip 0.25in
{\it
$^1${Department of Physics,
University of California, Santa Barbara, CA 93106, USA}\\
$^2${School of Physics and Astronomy,
University of Minnesota, Minneapolis, MN 55455, USA}
}\\
\vskip 0.5in
{\bf Abstract}
\end{center}
\baselineskip=18pt \noindent
We calculate the one-loop contribution to the bino annihilation rate due to
the process $\>\widetilde B \widetilde B\rightarrow Z^*\>$, which vanishes at
tree level.
\end{titlepage}
\baselineskip=18pt
{\newcommand{\la}{\mbox{\raisebox{-.6ex}{~$\stackrel{<}{\sim}$~}}}
{\newcommand{\ga}{\mbox{\raisebox{-.6ex}{~$\stackrel{>}{\sim}$~}}}
\def\mb{m_{\widetilde B}}
\def\ob{\Omega_{\widetilde B}}
\def\msf{m_{\widetilde F}}
\def\beq{\begin{equation}}
\def\eeq{\end{equation}}
It is very likely that some extension of the standard electroweak model
is needed.  In fact, such an extension is probably dictated by
cosmology in order to produce a baryon asymmetry and to provide a suitable
dark matter candidate. The former typically relies on grand unification
(though simpler extensions also work \cite{dol}), while the best motivated
extension which provides a cold dark matter candidate is the
minimal supersymmetric standard model (MSSM).
Because of an unbroken discrete symmetry, R-parity, the
lightest supersymmetric partner (LSP) is expected to be stable.
This makes the LSP a natural candidate
for the role of the non-baryonic dark matter.
In previous studies, the lightest neutralino, a linear combination
of the supersymmetric partners of the neutral $SU(2)$ gauge boson
(neutral wino), $U(1)$ hypercharge gauge boson (bino), and the two neutral
components of the Higgs doublets (higgsinos) has been suggested as the most
likely candidate for the LSP \cite{ehnos}.  In a large portion of parameter
space, the least massive eigenstate is a nearly pure bino \cite{os34}.

Neutralino annihilation rates have been calculated previously at tree level,
under a variety of assumptions and to various degrees of
accuracy \cite{ehnos,gr,os34,gkt,mos,n,dn,fmos,my}.  For neutralinos which are
nearly pure bino, the rate is dominated by creation of  fermion
pairs via the exchange of a sfermion, though many of the calculations
have included annihilation channels into Higgs scalars as well.  Since binos
are $SU(2)\times U(1)$ singlets, they do not couple directly to the $Z$.
Thus, s-channel annihilations through $Z$ exchange which are very important
for massive neutrino annihilations or higgsino co-annihilations \cite{gs,my},
do not occur  for pure binos. Of course in reality the LSP is never a ``pure"
bino, but rather always contains some small admixture of higgsinos as well.
Annihilations through $Z$ exchange due to this admixture make a small
contribution to the cross-section.
In this paper, we consider a correction to the bino annihilation rate from
the process $\widetilde B \widetilde B \rightarrow Z^*$, via the loop diagrams
shown in fig.~(1).  Generically, the corrections are found to be much larger
than the corrections due to the higgsino admixture.  However,
their contribution to the total annihilation cross-section remains small,
 on the order of a few percent or less over the physically allowed range of
parameters.  Recently one-loop corrections to the neutralino masses have been
computed \cite{corr}, and found to be very small (less than one per cent)
in the parameter range of interest here.  These corrections to the masses
would make similarly small corrections to the annihilation cross section,
and so are significantly smaller than those we compute here.

In general, the lightest neutralino mass eigenstate is a
linear combination of the gauginos $\widetilde W^3$ and $\widetilde B$,
and the higgsinos  $\widetilde H{}_1^0$ and $\widetilde H{}_2^0$:
\beq
\widetilde\chi^0 = \alpha \widetilde W^3 + \beta \widetilde B  + \gamma
\widetilde H_1^0 + \delta \widetilde H_2^0\;.
\eeq
In the pure bino region $(\beta\simeq 1)$, $m_{\widetilde\chi^0}
\simeq M_1$, where $M_1$ is
the supersymmetry breaking $U(1)$ gaugino mass. For our purposes here,
the supersymmetric parameter space can be described by one additional
parameter, $\varepsilon$, the supersymmetric Higgs mixing mass (often denoted
by $-\mu$). In the bino region, results are very insensitive to the ratio
of Higgs vevs, $\tan \beta$ ($\tan\beta$ should not be confused
with the $\beta$ in eq.~(1)).  In what follows we have chosen
$\tan \beta = 2$ and assumed a top quark mass of $160 \gev$.
We simplify by using GUT boundary conditions to relate $M_2$, the $SU(2)$
gaugino mass, to $M_1$ via $M_1 = {5\over3} M_2 \tan^2\theta_{\rm w}$.
{}From here on, we will assume that $\varepsilon \gg M_2$ so that
we are working in the nearly pure bino region and
write $m_{\widetilde\chi^0} = \mb$.

We begin by calculating the annihilation rate of almost pure binos
to fermion anti-fermion pairs $\bar F F$ via (1) t-channel sfermion exchange,
(2) $\widetilde B\widetilde B \to Z^*\to \bar F F$ via the loop diagrams of
fig.~(1), and (3) $\widetilde B\widetilde B \to Z^*\to \bar F F$ via the
small but non-zero admixture of $\widetilde H_1^0$ and $\widetilde H_2^0$ in
$\widetilde\chi^0$ (see eq.~(1)).

The relevant interaction terms are
\begin{eqnarray}
{\cal L} &=& -\;{g_2\over{4\cos\theta_{\rm w}}}\;
             Z_\mu\left(\overline{\widetilde H}{}_1^0
                        \gamma^\mu\gamma_5\widetilde H_1^0
    -\overline{\widetilde H}{}_2^0 \gamma^\mu\gamma_5\widetilde H_2^0\right)
             + \sum_f \biggl\{{g_1\over\sqrt2}
               \bar f\,\bigl(Y_{f_R}\widetilde f_R P_-
                           + Y_{f_L}\widetilde f_L P_+\bigr) \widetilde B
             +{\rm h.c.} \nonumber \\
&{}& \qquad
             +\;{g_2\over\cos\theta_{\rm w}}\;
              Z_\mu\Bigl[\,\bar f \gamma^\mu
  (C_V^f - C_A^f \gamma_5) f  + i\, (C_V^f + C_A^f)(\widetilde f^*\partial^\mu
\widetilde f - \widetilde f\partial^\mu\widetilde f^*)\Bigr]\biggr\}
\end{eqnarray}
where $P_\pm=(1\pm\gamma_5)/2,\,
Y_{f_R} = 2 Q_f,\, Y_{f_L} = 2 (Q_f - T_3^f),\, C_V^f = -T_3^f/2 +
Q_f \sin^2\theta_{\rm w}\, , $ and $C_A^f = -T_3^f/2$.  The fermion charge is
denoted by $Q_f$,
and $T_3^f = +1/2$ for up-type quarks and neutrinos, and $-1/2$ for down-type
quarks and charged leptons.  The sum runs over all fermions.  We neglect
the effect of sfermion mixing, which is treated in detail in ref.~\cite{fmos}.

To derive a thermally averaged
cross section, we make use of the technique of ref.~\cite{swo}.  We expand
$\langle\sigma v_{\rm rel}\rangle$ in a Taylor expansion in powers of
$x = T/m_{\widetilde B}$:
\beq
\langle \sigma v_{\rm rel } \rangle = a + b x + O(x^2)\;.
\eeq
The coefficients $a$ and $b$ are given by
\begin{eqnarray}
a &=& \sum_F v_F \tilde a_F
\nonumber \\
b &=& \sum_F v_F \left[ \tilde b_F + \left( -3 + {3 m_F^2 \over 4 v_F^2
    m_{\widetilde B}^2 } \right) \tilde a_F \right]
\end{eqnarray}
where the subscript $F$ specifies the final state fermions,
$ \tilde a_F $ and $ \tilde b_F $ are computed from the expansion of the
matrix element squared in powers of $p$, the incoming bino momentum, and
$v_F = (1 - m_F^2/m_{\widetilde B}^2)^{1/2}$
is a factor from the phase space integrals.
In terms of the squared reduced transition matrix element $|{\cal T}|^2$,
$\tilde a_F$ can be written simply as,
\beq
\tilde a_F = {1\over 32\pi} {1\over \mb^2} |{\cal T}_F(p=0)|^2
\eeq
Denoting the reduced transition amplitudes for the processes (1), (2) and
(3) listed above by ${\cal T}_{\rm sfex},\, {\cal T}_{\rm loop}$ and
${\cal T}_{\rm mix}$ respectively, we have
\beq
|{\cal T}_F| = |{\cal T}_{\rm sfex} +{\cal T}_{\rm loop} +
                  {\cal T}_{\rm mix}|\;,
\eeq
where
\begin{eqnarray}
{\cal T}_{\rm sfex}(p=0) &=& {g_1^2\over 2} \;(Y_{F_L}^2 +
   Y_{F_R}^2)\; {m_F\mb \over \mb^2 - m_F^2 + m_{\widetilde F}^2} \\
\noalign{\medskip}
{\cal T}_{\rm mix}(p=0) &=&
  -2\,\left(g_2\over\cos\theta_{\rm w}\right)^2\,C^F_A\,{m_F\mb\over m_Z^2}\,
        (\gamma^2 - \delta^2) \\
\noalign{\medskip}
{\cal T}_{\rm loop}(p=0) &=&
 +\,{2g_2\over\cos\theta_{\rm w}}\,C_A^F\,{m_F\mb\over m_Z^2}\,
  {g_1^3\over \sin\theta_{\rm w}}\,{1\over 32 \pi^2}\times\cr
\noalign{\smallskip}
&{}&\mskip 25mu
\sum_f \biggl\{Y_{f_L}^2 \Bigl[2(T_3^f - Q_f\sin^2\theta_{\rm w})-
T_3^f I_f\Bigr]+Y_{f_R}^2\Bigl[2Q_f\sin^2\theta_{\rm w}-T_3^f I_f
    \Bigr]\biggr\} \cr
\noalign{\medskip}
&=&  -\,{2g_2\over\cos\theta_{\rm w}}\,C_A^F\,{m_F\mb\over m_Z^2}\,
  {g_1^3\over \sin\theta_{\rm w}}\,{1\over 32 \pi^2}
\sum_f T_3^f\,(Y_{f_L}^2 + Y_{f_R}^2)\,I_f\biggr\}
\end{eqnarray}
and where
\beq
I_f = \int_0^1 dr\int_{-r}^r ds\;{m_f^2\over
(m_f^2 - m_{\widetilde f}^2 - m_{\widetilde B}^2)\,r + m_{\widetilde B}^2\,s^2
+ m_{\widetilde f}^2 + i\varepsilon}\;.
\eeq
Here $C^F_A,\, Y_{F_L},\, Y_{F_R}$ are the quantum numbers
of the final state fermion, and $m_F$ and $m_{\widetilde F}$ are
the masses of the final state fermion and its supersymmetric partner.
These expressions for the amplitudes do not include spin dependence, but
we note that they will vanish if either the binos or fermions
are spin parallel. In the other cases the amplitudes are antisymmetric
under exchange of spins in the incoming or outgoing states. So their
relative signs remain the same allowing us to directly sum and average over
spin states giving eq.~(5). To compute $\tilde b_F$, where we need to
consider $p\ne 0$, this simple spin dependence no longer holds
and we use standard trace techniques to calculate the squared transition
matrix element.

To compare the relative effect of the loop and the higgsino mixing
diagrams on the dominant sfermion exchange process, we need to
find the $\tilde a_F$ and $\tilde b_F$ coefficients for 1) the
sfermion exchange alone, 2) the sfermion exchange plus the loop
diagram and 3) the sfermion exchange plus the higgsino mixing
diagram. The $\tilde a_F$ for each combination can be read off
from eq.~(5) by redefining
${{\cal T}_F}$ to contain different combinations of diagrams,
but the $\tilde b_F$ will need to be separately
computed in each case. The expressions are too complicated to reproduce
here.

We now discuss the effect of this loop correction on the relic density and
detectability of binos.
In fig.~(2), we show a contour plot of $\ob h^2$ as a function of $\mb$ and
the common sfermion mass $m_{\widetilde f}=m_{\widetilde F}$,
assuming that binos annihilate to fermions
only through the sfermion exchange process (1).
$\ob$ is the mass density of the binos
in units of the critical density, and $h$ is the Hubble parameter in units
of 100 km s$^{-1}$ Mpc$^{-1}$.  The region where $\msf>\mb$ is excluded,
since here the LSP would be a sfermion rather than the bino.
Notice the effect of the top threshold
at $\mb = 160\gev$.  Because the cross-section is p-wave suppressed, we have
$a_F \propto (m_F^2/\mb^2)$, and so as we cross
below $\mb = m_{\rm top}$ the relic density rises sharply.
Notice also that $\ob h^2 \le 1$ requires $\msf\la 500\gev$, while
$\ob h^2 \le 1/4$ requires $\msf\la 350\gev$. Keep in mind that these
are approximate values for $\ob h^2$ as only the dominant part of the
cross-section was used here, though the corrections are very small.

In fig.~(3), we compare the effect of including the loop diagrams (2)
 on the relic density of binos $\ob h^2$ to the effect of including the
higgsino mixing diagram (3). As $\varepsilon$ gets larger,
the admixture of higgsino
in  $\widetilde\chi^0$ gets smaller, and so does the prefactor $(\gamma^2 -
\delta^2)$ in eq.~(8) (and in the corresponding $p\ne 0$ expression).
In the exactly pure bino limit, corresponding to $\varepsilon\to\infty$
at fixed $M_1$, the process $\widetilde\chi^0\widetilde\chi^0\to Z$ is
forbidden, as mentioned above.  For each point in the
$\mb-\msf$ plane we find the value of $\varepsilon$
at which the magnitude of the loop correction
is equal to that of the higgsino mixing correction. We then plot
several contours of these $\varepsilon$ values.
For each $\varepsilon$, it is the region to
the left of the corresponding contour where the loop effect is
greater.  In the parameter range of interest, we find that
the loop correction to $\ob h^2$ vanishes near the line $\msf = \mb$,
which is why  $\varepsilon$ tends toward large values there.
For the  moderate values of $\msf$ represented in fig.~(3), we see that the
effect of the loop is larger than the effect of higgsino mixing for large
ranges in $\mb$ in the pure bino regime.  Again, we exclude the region where
$\mb > \msf$. We also exclude the region below the top threshold where both
diagrams have a negligible effect on $\ob h^2$.

Although we have found that for a large portion of the parameter
space the loop correction is larger than the commonly used
correction due to a higgsino admixture, the net effect on the
annihilation cross-section and ultimately on $\ob h^2$ is small.
For all values of
$\msf$ and $M_2$ consistent with $\ob h^2 \le 1$, the effect is at  most
a few percent, and in the allowed region, the effect predominantly
decreases the relic density.
Below the top threshold, the effect of the loop diagrams drops by
a factor of $\sim\!300$ due to the accumulation of a number of unrelated
factors.  (This is also true for the higgsino mixing diagram.)
Of course in the very thin slice of parameter
space near $\mb = m_Z/2$, the annihilation cross-section will
be greatly enhanced, and the effect of the
loop diagrams will be significant.  Because this effect is limited
to a small region of parameter space we do not consider it further.
We note only that in such regions the expansion of the thermal
average should be done as explained in ref.~\cite{gs}.

In addition to the effect of the loop correction on the relic density,
it is also of interest to compute the effect on the zero temperature
annihilation cross-section ($p=0$) which governs annihilations in the galactic
halo. Indeed, galactic halo LSP annihilation products have been
frequently discussed as a potential signatures for dark matter \cite{gamma}.

In fig.~(4), we show the effect of the loop diagrams on the zero-temperature
cross-section,  $\langle \sigma v_{\rm rel} \rangle$, at $\varepsilon=\infty$.
The contours show the percentage change in $a$ due to the inclusion of the loop
correction diagrams (2). The effect is larger than on the relic density,
but in the physically allowed region the effect is still
$\la 11\%$.  We again see the top threshold at $\mb = 160 \gev$.  Notice
that below the top threshold, the sign of the effect changes.  At
zero incoming bino momentum, the coupling of the $Z$ to the final state
fermions is proportional to $T_3^F$, as seen in eq.~(9).
Above the top threshold, zero temperature annihilation
is primarily to tops, while below the top threshold it is primarily to
$b$'s and $\tau$'s.  Thus ${\cal T}_{\rm loop}(p=0)$ switches sign, while
${\cal T}_{\rm sfex}(p=0)$ does not.

Finally in fig.~(5), we compare the effect of including the loop diagrams (2)
on the zero-temperature cross-section $a$
to the effect of including the higgsino mixing diagram (3).
Calling the change in $a$ due to the loop diagrams $\Delta_{\rm loop}a$
and the change in $a$ due to the higgsino mixture $\Delta_{\rm hm}a$,
we plot $\Delta_{\rm loop}a/\Delta_{\rm hm}a$ as a function of
$\varepsilon$ and $\mb$ using $\msf = 500$ GeV.
As $\varepsilon$ becomes large, the effect of the loop correction greatly
exceeds  the effect of higgsino mixing except very near to the value of
$\mb$ at which the loop's effect on the cross section vanishes.

To summarize, we have calculated the effect of the loop diagram of
fig.~(1) on the annihilation cross-section for binos.  We find that
away from the $Z$-pole, the effect of the additional diagram is to
change the relic density of binos $\ob h^2$ by at most a few percent. This
is true over the physically allowed range of common sfermion mass $\msf$ and
bino mass $\mb$, though it is larger than  contributions
from the direct annihilation
to $Z$'s via the small but non-zero admixture of higgsinos in
$\widetilde\chi^0$, typically included in this type of calculation
for a wide range in values of the higgsino mixing mass $\varepsilon$.
Near the $Z$-pole, we expect the effect of loop corrections to be
considerably larger.  The effect of the loop diagram on the zero-temperature
cross-section is larger and may be as large as $11\%$.

\bigskip

\vbox{
\noindent{ {\bf Acknowledgements} } \\
\noindent  This work was supported in part by DOE grant DE--FG02--94ER--40823
and NSF grant PHY--91--16964.  The work of KAO was in addition supported by a
Presidential Young Investigator Award.
}}
}

\newpage
\noindent{\bf{Figure Captions}}

\vskip.3truein

\begin{itemize}
\def\mb{m_{\widetilde B}}
\def\ob{\Omega_{\widetilde B}}
\def\msf{m_{\widetilde F}}

\item[]
\begin{enumerate}
\item[]
\begin{enumerate}

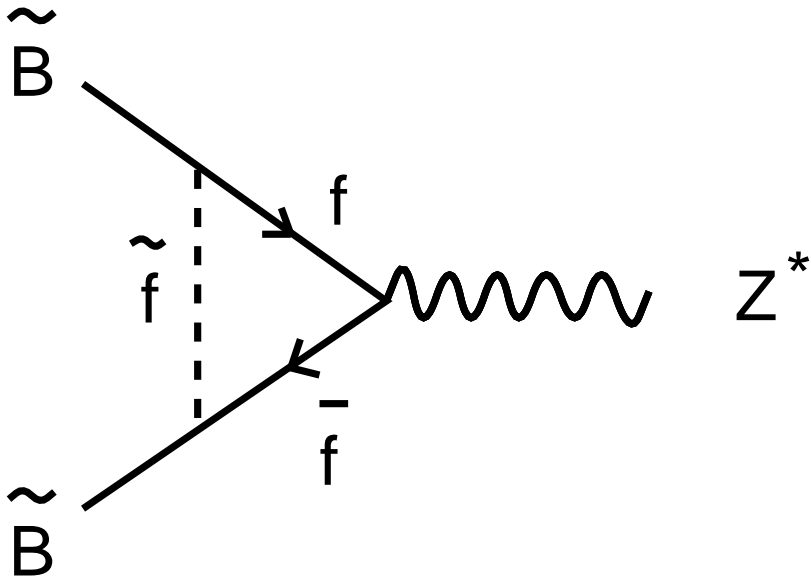
\item[Fig.~1)] Feynman diagrams for loop corrections to
$\widetilde B \widetilde B \to Z^*$.

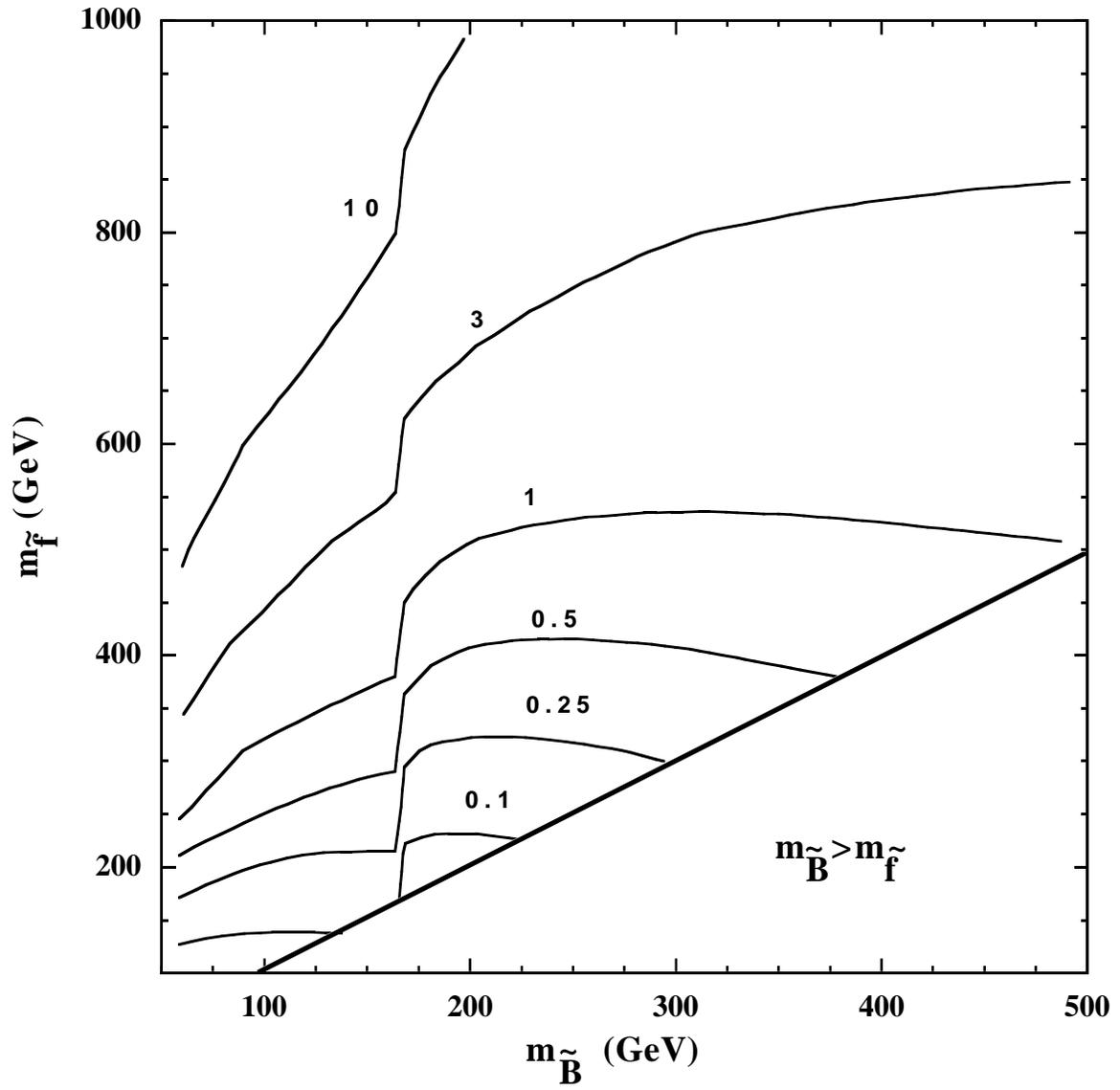
\item[Fig.~2)] Relic density $\ob h^2$ of binos, assuming bino annihilation
only due to sfermion exchange.  We take $\tan\beta=2, m_{\rm top}=160\gev$,
and a common sfermion mass $m_{\widetilde F}=m_{\widetilde f}$.

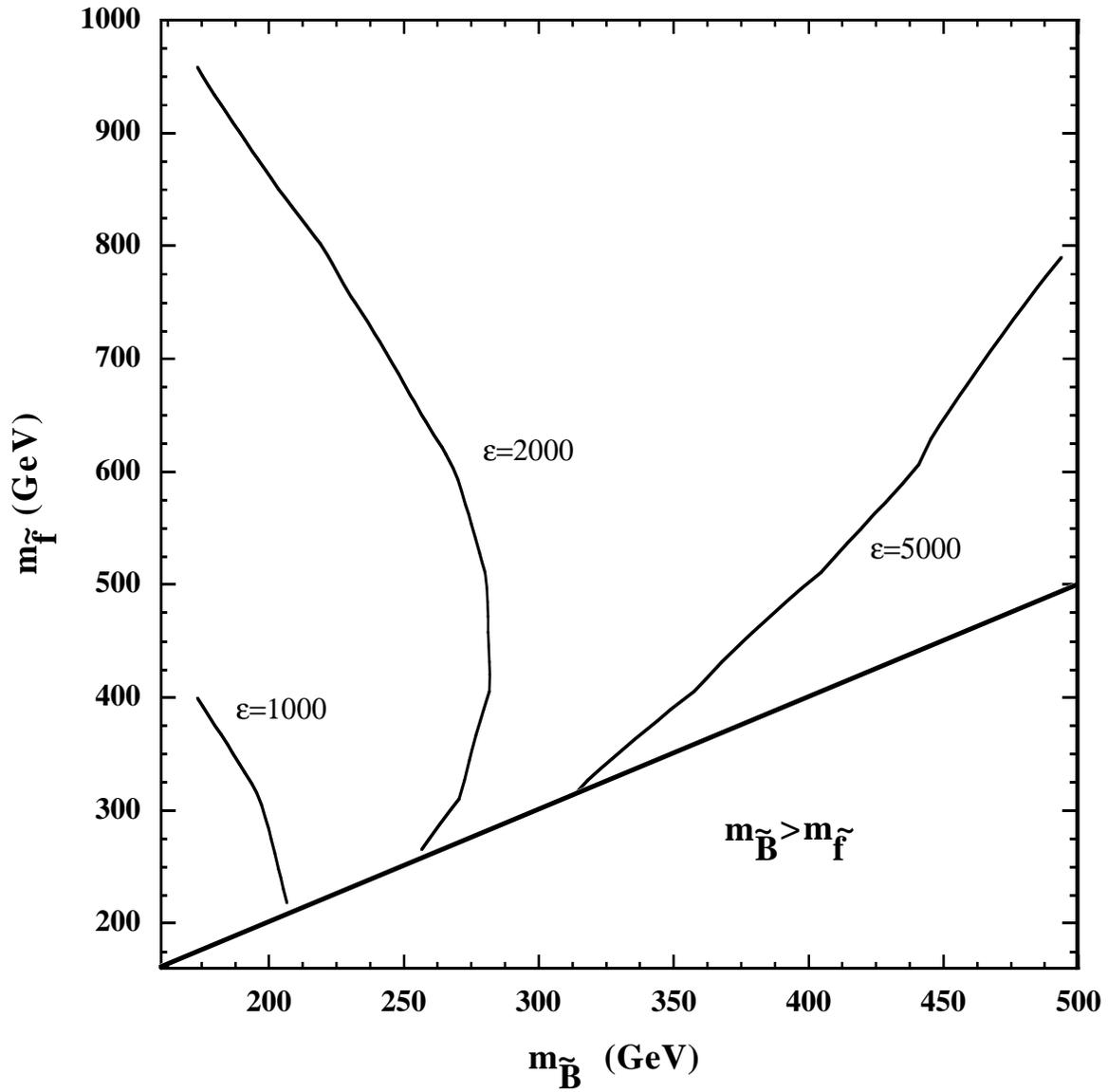
\item[Fig.~3)] A comparison of the effects of the loop diagrams and the
higgsino mixing diagram on the relic density of binos $\ob h^2$, for several
values of $\varepsilon$.  In the area to the left of each curve labelled by
$\varepsilon$, the loop correction has a greater
effect on $\ob h^2$ than the higgsino mixing correction.

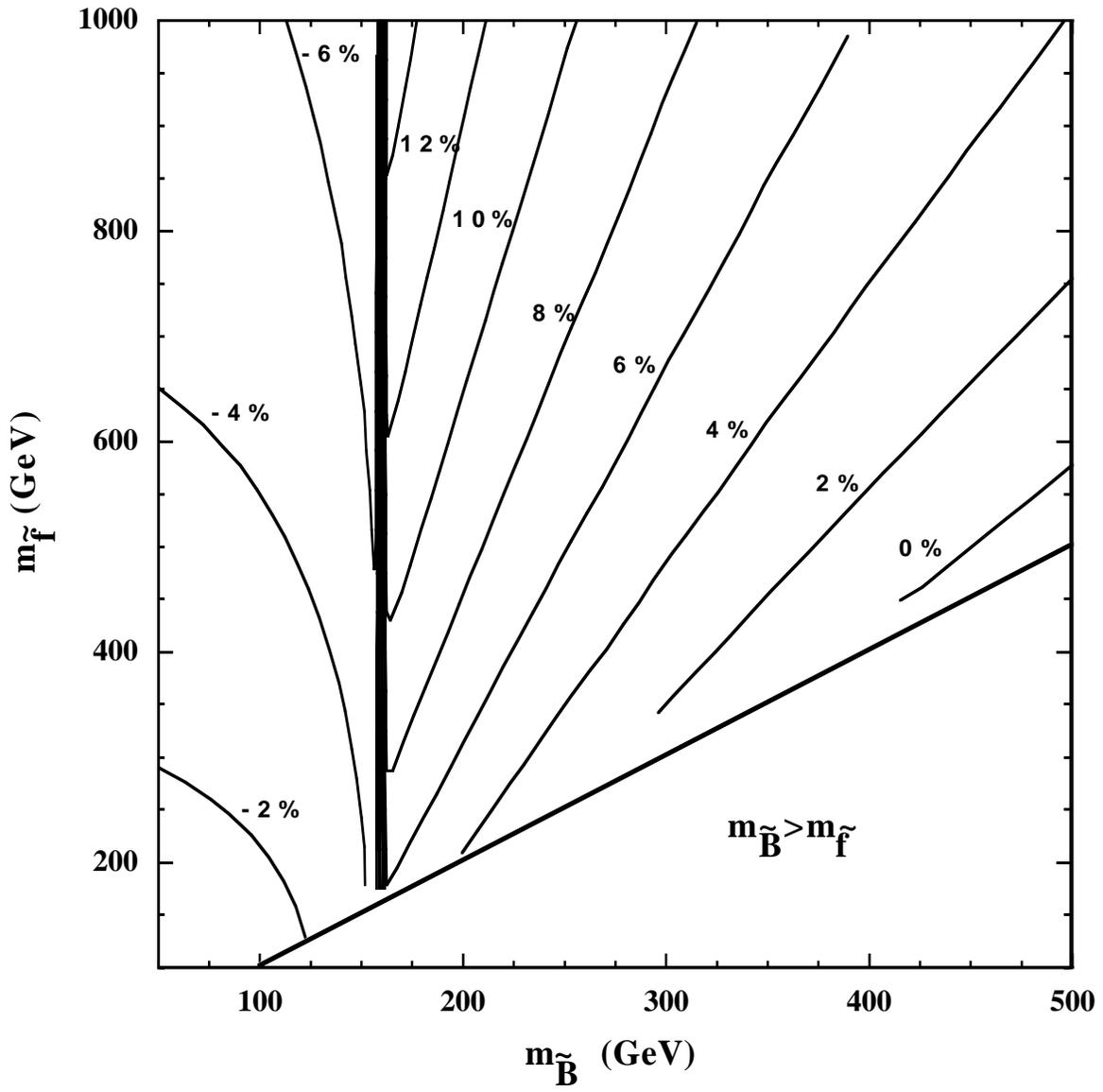
\item[Fig.~4)] The effect of the loop diagrams on the zero-temperature
cross-section.  Contours label the percentage change in $a$ due to the
inclusion of the loop correction.

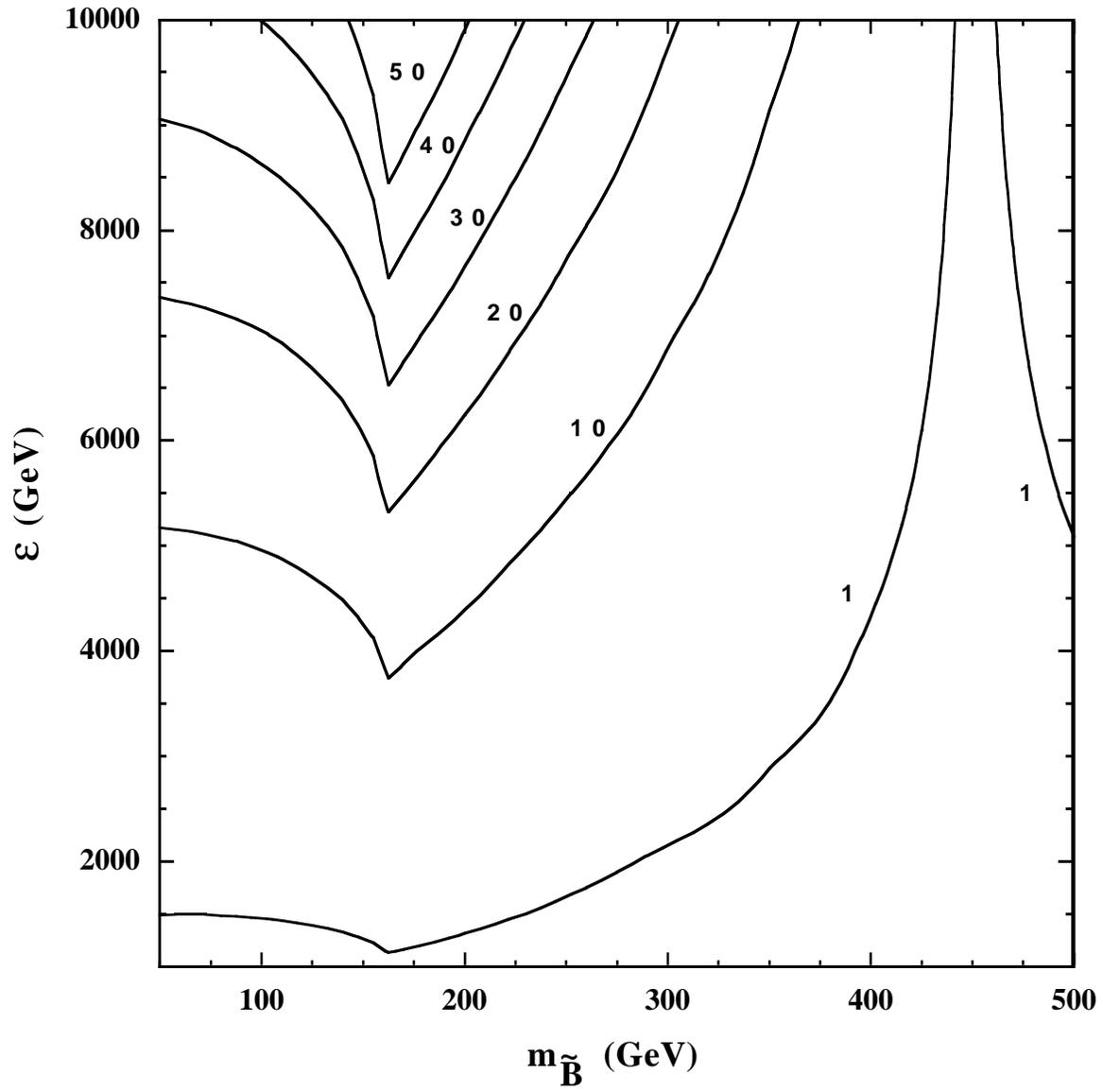
\item[Fig~5)] A comparison of the effects of the loop diagrams  and the
higgsino mixing diagram  on the zero-temperature cross-section.

\end{enumerate}
\end{enumerate}
\end{itemize}
\end{document}